\documentclass[12pt]{book}

\usepackage[dvips]{graphicx,color}
\usepackage{makeidx,view}

\makeauthorindex
\makeindex

\begin{document}

\BookTitle{\itshape The Universe Viewed in Gamma-Rays}
\CopyRight{\copyright 2002 by Universal Academy Press, Inc.}
\pagenumbering{arabic}

\chapter{
Recent Results from the VERITAS Collaboration }

\author{F. Krennrich$^{1}$, I.H.~Bond$^{2}$,
S.M. Bradbury$^{2}$, J.H. Buckley$^{3}$, D.A.~Carter-Lewis$^{1}$,
W.~Cui$^{4}$, I.~de la Calle Perez$^{2}$, A.~Falcone$^{4}$,
D.J.~Fegan$^{5}$, S.J.~Fegan$^{6, 15}$, J.P.~Finley$^{4}$,
J.A.~Gaidos$^{4}$, K.~Gibbs$^{6}$, G.H.~Gillanders$^{7}$,
T.A.~Hall$^{1, 13}$, A.M. Hillas$^{2}$, J.~Holder$^{2}$,
D.~Horan$^{6}$, M.~Jordan$^{3}$, M.~Kertzman$^{8}$,
D.~Kieda$^{9}$, J.~Kildea$^{5}$, J.~Knapp$^{2}$, K.~Kosack$^{3}$,
M.J.~Lang$^{7}$, S.~LeBohec$^{1}$, P.~Moriarty$^{10}$,  D.
M\"uller$^{11}$, R.A.~Ong$^{12}$, R.~Pallassini$^{2}$,
D.~Petry$^{1}$, J.~Quinn$^{5}$ , P.T. Reynolds$^{14}$,
H.J.~Rose$^{2}$, G.H. Sembroski$^{4}$, S.P.~Swordy$^{11}$, V.V.
Vassiliev$^{9}$, S.P.~Wakely$^{11}$,T.C.~Weekes$^{6}$}

\author{$^{1}$Physics \& Astronomy Department,
Iowa State University, Ames, IA 50011, USA \\ $^{2}$Department of
Physics, University of Leeds, Leeds, LS2 9JT, Yorkshire, UK\\
$^{3}$Department of Physics, Washington University, St.~Louis, MO
63130\\
$^{4}$Department of Physics, Purdue University, West Lafayette, IN
47907\\ $^{5}$Physics Department, National University of Ireland,
Belfield, Dublin 4, Ireland \\
$^{6}$F. Lawrence Whipple Observatory, Harvard-Smithsonian CfA,
Amado, AZ 85645 \\
$^{7}$Physics Department, National University of Ireland, Galway,
Ireland \\
$^{8}$Physics Department, De Pauw University, Greencastle, IN,
46135 \\
$^{9}$High Energy Astrophysics Inst., University of Utah, Salt
Lake City, UT 84112 \\
$^{10}$School of Science, Galway-Mayo Institute of Technology,
Galway, Ireland \\
$^{11}$Enrico Fermi Institute, University of Chicago, Chicago, IL
60637 \\
$^{12}$Department of Physics, University of California, Los
Angeles, CA 90095
\\
$^{13}$Department of Physics, University of Arkansas at Little
Rock, AR 72204}


%
\AuthorContents{R.\ Enomoto M.\ Mori, and S.\ Yanagita} 

\section*{Abstract}

A decade after the discovery of TeV $\gamma$-rays from the blazar
Mrk~421 (Punch et al. 1992),  the list of TeV blazars has
increased to five BL Lac objects: Mrk~421 (Punch et al. 1992;
Petry et al. 1996; Piron et al. 2001), Mrk~501 (Quinn et al. 1996;
Aharonian et al. 1999; Djannati-Atai et al. 1999), 1ES2344+514
(Catanese et al. 1998), H1426+428 (Horan et al. 2000, 2002;
Aharonian et al. 2002; Djannati-Atai et al. 2002) and 1ES1959+650
(Nishiyama et al. 1999; Konopelko et al. 2002; Holder et al.
2002). In this paper we report results from recent observations of
Mrk~421, H1426+428 and 1ES1959+650 using the Whipple Observatory
10~m telescope.

\section{Introduction}

The scientific interest in TeV blazars is manifold and ranges from
classification studies of the AGN (active galactic nucleus)
phenomenon and detailed studies of Doppler-boosted relativistic
jets, to particle acceleration and $\gamma$-ray emission models
near the central massive object. In addition, $\gamma$-ray beams
traversing intergalatic space can be used to constrain and measure
the diffuse extragalactic background light (EBL) through
$\gamma\gamma \rightarrow e^{+}e^{-}$ absorption. In this paper we
present recent results relevant to those topics.

\bigskip

\section{Broad Studies: Mrk~501, Mrk~421, 1ES2344+514, H1426+428 \& 1ES1959+650}

 Various search strategies have been explored to build a
catalogue of TeV blazars. The discovery of Mrk~421 as a TeV blazar
resulted from a survey of the first EGRET catalogue of blazars
(Fichtel et al. 1994), however, none of the other EGRET blazars
have been detected.  The focus shifted then to nearby BL Lac
objects which led to the discovery of Mrk~501 as a $\gamma$-ray
source (not in first EGRET catalogue), indicating that TeV blazars
require search strategies independent from GeV sources.  A
striking feature in the multiwavelength spectrum of Mrk~501 is the
synchrotron peak at an energy above 100~keV (Catanese et al. 1997;
Pian et al. 1998) and a second peak at several hundred GeV
(Samuelson et al. 1998). This provided the impetus for a refined
search strategy (Catanese \& Weekes 1999) selecting BL Lacs that
are X-ray bright with spectra extending to hard X-rays. A
systematic study that is based on a selection of X-ray and radio
bright BL Lacs from X-ray and radio catalogues (Costamante \&
Ghisellini 2002) yielded a number of good candidates for TeV
emitting BL Lacs, with H1426+428 and 1ES1959+650 among them. In
fact H1426+428 was recently detected by several groups (Horan et
al. 2002; Aharonian et al. 2002; Djannati-Atai et al. 2002)
establishing the most distant TeV blazar that has been confirmed
by several groups, with a redshift of z=0.129.

Another recent detection from the list of X-ray selected TeV
candidate BL Lacs (Costamante \& Ghisellini 2002) is 1ES1959+650.
Observations between May 16th and July 8th 2002 yielded a strong
detection (Dowdall et al. 2002; Holder et al. 2002) showing
$\gamma$-ray flux states between 0.5 and 5~times the Crab Nebula
flux. Due to its slightly larger redshift (z=0.048) in comparison
to  Mrk~421 and Mrk~501 but smaller redshift than that of
H1426+428, spectral studies of 1ES1959+650 will provide important
additional information about the EBL (see also Holder et al. in
these proceedings). In summary, the catalogue of TeV blazars is
steadily increasing, indicating that they are becoming an
important class of extragalactic objects for high energy
astrophysics at TeV energies.

\begin{figure}[t]
  \begin{center}
    \includegraphics[height=20pc]{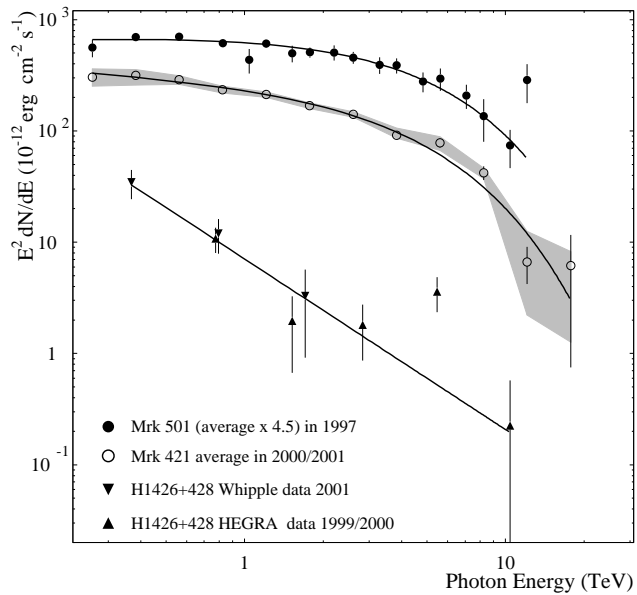}
  \end{center}
  \caption{The average energy spectrum of Mrk~501 (filled circles) in 1997 (Samuelson et al. 1998) is shown
            in comparison to the average spectrum of Mrk~421 (empty circles) in 2000/2001 (Krennrich et al. 2001).
            Furthermore, the combined energy
             spectrum of H1426+428 using data from the HEGRA collaboration (Aharonian et al. 2002) is shown by
             the upward triangles,  and data from the Whipple collaboration (Petry et al. 2002) is shown by the
              downward pointing triangles.}
\end{figure}

\section{Spectral cutoffs and variability: EBL}
Spectroscopic information from $\gamma$-ray blazars is the key to
extracting the physics related to acceleration and emission
processes in the relativistic jet.   In addition, spectra may
exhibit imprints from absorption effects of $\gamma$-rays due to
radiation fields in the vicinity of the AGN, or attenuation of the
$\gamma$-ray beam by the EBL. Of particular interest has been the
search for signatures of $\gamma$-ray absorption by the EBL in the
energy spectra of Mrk~501 and Mrk~421, since high flaring states
in 1997 and 2001, respectively, provided excellent statistics.

Figure~1 shows the average energy spectra of Mrk~501 and Mrk~421
for these high flaring states with flux levels reaching several
Crab. Mrk~421 and Mrk~501 have similar redshifts and exhibit a
cutoff in their energy spectra at $\rm 4.3 \pm 0.3_{stat}$~TeV and
$\rm 4.6 \pm 0.8_{stat}$~TeV, respectively. The consistent cutoff
energy of approximately 4~TeV for both sources might suggest, that
it is of common origin.  The similarity of their distance  to
Earth makes absorption due to the EBL a viable possibility. The
cutoffs could also be caused by a falling Klein-Nishina
cross-section in an inverse-Compton picture.  Other possibilities
include absorption by radiation fields near the $\gamma$-ray
source or a terminating particle energy distribution in the jet.
However, since the historical synchrotron peak energies observed
for both sources differ by almost two orders of magnitude, the
latter scenario seems less likely.

To decide whether or not the observed cutoff is indeed due to the
EBL, measurements of TeV spectra for sources at different
redshifts are necessary. Although those studies are in their
infancy, the energy spectrum of one of the recently detected
sources with a larger redshift provides some important clues. Also
shown in Figure~1 is the energy spectrum of H1426+428 between
400~GeV and 10~TeV, which can be well described by a power law
with a differential flux following $\rm dN/dE \propto E^{-3.55 \pm
0.27}$ (Petry et al. 2002).   This makes H1426+428 the TeV blazar
with the softest spectrum above 400~GeV, despite the fact that
H1426+428 seems to reach at times a synchrotron peak energy up to
$\rm E > 100$~keV (Costamante et al. 2001). H1426+428 is the most
distant BL Lac seen at TeV energies and theoretical studies of the
EBL (Stecker \& Salamon 1992; Primack 2002) suggest that its TeV
$\gamma$-ray spectrum may be substantially attenuated through pair
production off the EBL. Although the steep energy spectrum of
H1426+428 concurs well with this picture, studies that could
reveal the detailed shape of an attenuated spectrum require better
statistics. The data published so far are consistent with a power
law. Nevertheless, future measurements with better statistics
should be useful for constraining  the EBL density.

A promising tool in the quest for understanding the origin of
spectral cutoff features is the study of spectral variability.
Flaring activity of Mrk~421 in 2001 yielded $\rm \approx 23,000$
$\gamma$-rays at energies $\rm E > 300$~GeV providing the
statistics needed to carry out studies of spectral variability
over an order of magnitude of flux levels (set I - set VIII).
Figure 2 shows energy spectra for the 2000/2001 data of Mrk~421
binned into different flux levels. Spectral hardening with
increasing flux levels is apparent from the data.  Figure 3 shows
the spectral index plotted versus flux (units of Crab), providing
the first unequivocal evidence that TeV $\gamma$-ray spectra do
indeed vary, placing an important observational constraint on
models of particle acceleration and $\gamma$-ray emission in jets
of AGNs.  Studies of the variability of the cutoff energy should
also provide important clues about the cause of the cutoff (for
further discussion see Krennrich et al. 2002).

In summary, we have extended the catalogue of TeV blazars to 5~ BL
Lac objects.  The study of their energy spectra demonstrate that
spectroscopy of TeV blazars is becoming a promising means to
constrain the emission mechanism in their jets.  In addition, an
indirect measurement of the EBL between $\rm 0.1$ and $\rm 30
\mu$m is coming into reach, which will most likely be achieved by
the next generation ground-based $\gamma$-ray telescopes CANGAROO,
HESS, MAGIC and VERITAS.

\begin{figure}[t]
  \begin{center}
    \includegraphics[height=20pc]{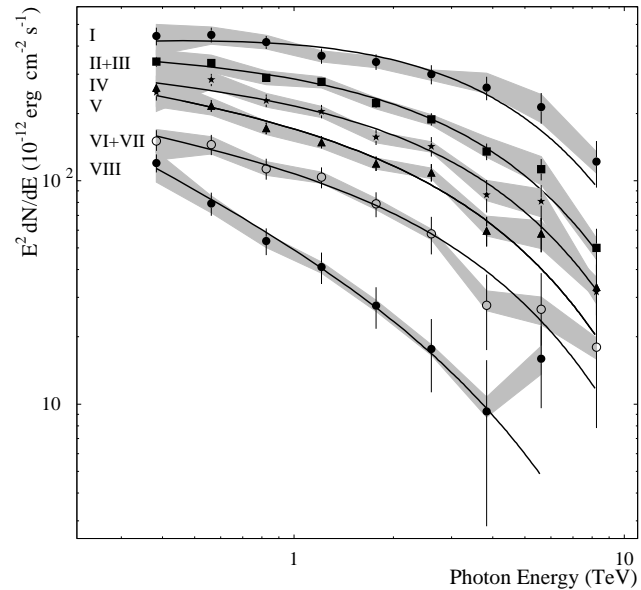}
  \end{center}
  \caption{The energy spectra of Mrk~421 averaged over 5 months are shown for different flux levels ranging
            from 1.3 - 10.5~Crab (Krennrich et al. 2002).}
\end{figure}

\begin{figure}[t]
  \begin{center}
    \includegraphics[height=20pc]{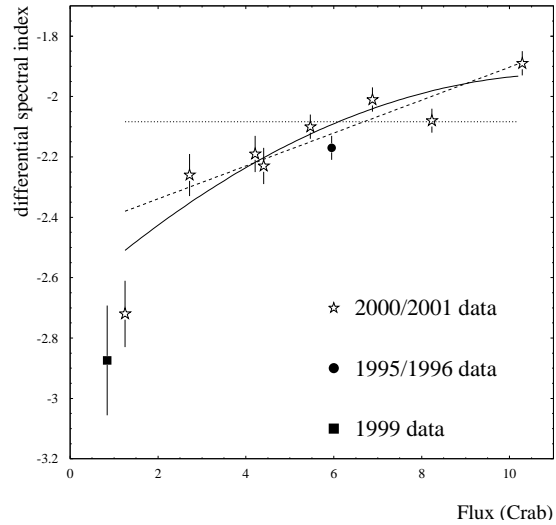}
  \end{center}
  \caption{The stars show the spectral index of Mrk~421 as a function of the flux (in units of Crab)
           for the 2000/2001 data, the circle shows a data point for Whipple 1995/96 data
            (Krennrich et al. 1999) and the square represents data taken during 1999 May-June.}
\end{figure}

\section{References}


\re 1.\ Aharonian, F.A.\ et al.,\ 1999, A\&A, 349, 11

\re 2.\ Aharonian, F.A.\ et al.,\ 2002, astro-ph/0202072

\re 3.\ Catanese M.\ et al.\ 1997, ApJ, 487, L143

\re 4.\ Catanese M.\ et al.\ 1998, ApJ, 501, 616

\re 5.\ Catanese M.\ \& Weekes, T.C. 1999, PASP, 111, 1193

\re 6.\ Costamante, L. \& Ghisellini, G. \ 2002, A\&A, 384, 56

\re 7.\ Costamante, L. \ et al. \ 2001, A\&A, 371, 512

\re 8.\ Djannati-Atai,  A.\ et al.,\ 1999, A\&A , 350, 17

\re 9.\ Djannati-Atai,  A.\ et al.,\ 2002, astro-ph/0207618

\re 10.\ Dowdall, C.\ et al., \ 2002, IAUC, 7903

\re 11.\ Fichtel, C.A.\ et al.,\ 1994, ApJS, 94, 551

\re 12.\ Konopelko, A.\ et al.,\ 2002,  HEAD Meeting, Albuquerque,
B17, 2002

\re 13.\ Krennrich, F.\ et al.,\ 1999, ApJ, 511, 149

\re 14.\ Krennrich, F.\ et al.,\ 2001, ApJ, 560, L45

\re 15.\ Krennrich, F.\ et al.,\ 2002, ApJ, 575, L9

\re 16.\ Nishiyama,  T.\ et al.,\ 1999, Proc. of 26th ICRC, Salt
Lake City, 1999, 3, 370

\re 17.\ Holder,  J.\ et al.,\ 2002, submitted to ApJL

\re 18.\ Horan, D. et al.,\ 2000, Head Meeting (Honolulu, Hawaii),
No. 32, 05.03

\re 19.\ Horan, D. et al.,\ 2002, ApJ, 571, 753

\re 20.\ Pian, E.\ et al.,\  1998, ApJ, 492, L17

\re 21.\ Punch  M.\ et al.,\  1992, Nature 358, 477

\re 22.\ Petry D.\ et al.,\  1996, A\&A 311, L13

\re 23.\ Petry D.\ et al.,\  2002, ApJ, 580, 104

\re 24.\ Piron F.\ et al.,\ 2001, A\&A , 374, 895

\re 25.\ Primack J.R.\ 2002, astro-ph/0201119

\re 26.\ Quinn J.\ et al.\ 1996, ApJ, 456, L83

\re 27.\ Samuelson, F.\ et al.\ 1998, ApJ, 501, L17

\re 28.\ Stecker, F.\, De Jager, O.C. \& Salamon, M.H. 1992, ApJ,
390, L49



%

 \chapter*{ Entry Form for the Proceedings }

\section{Recent Results from the VERITAS Collaboration}


\endofpaper
\end{document}